\documentclass[11pt]{article}
\usepackage{dsfont, amssymb,amsmath,amscd,latexsym, amsthm, amsxtra,amsfonts, mathrsfs}
\usepackage{graphicx}
\usepackage[all]{xy}
\begin{document}
\baselineskip 18pt
\title{{ { Strong Spatial Mixing for Binary Markov Random Fields}}}
\author{Jinshan \ Zhang\thanks{Corresponding author:
zjs02@mails.tsinghua.edu.cn} Heng Liang \ and Fengshan Bai
\\ \small Department of Mathematical Sciences,
\  Tsinghua University,\\
\small Beijing 100084,  China }
\date{}
\maketitle

{\small\begin{center}\textbf{Abstract}\end{center}

Gibbs distribution of  binary Markov random fields on a sparse on
average graph is considered in this paper. The
 strong spatial mixing is proved under  the condition that the `external field' is
uniformly large or small. Such condition on `external field' is
meaningful in physics.\\\\
\text{}}\\\\\\
\textbf{\large{ 1. Introduction}}

Strong spatial mixing property of Gibbs measures is  very important
 in statistical physics. It roughly says that if there
is a modification (or perturbation) on  the boundary conditions, its
influence to the Gibbs measure of a single vertex  decays
exponentially fast as the distance to the $support$ $of$ $the$
$perturbation$ (the set of vertices, whose spins are changed)
becomes large. In the classic literatures, it is also required that
the support of the perturbation has to be a single
vertex\cite{BMP99}. Weitz considers the support of perturbation to
be a set of vertices of arbitrary size. This generalized  definition
 is equivalent to the one in \cite{BMP99}
when the graph grows sub-exponentially (e.g. integer lattices). In
fact, the definition by Weitz has much wider application. For
example, it provides a natural algorithm to calculate the partition
function of Gibbs measures if the strong spatial mixing
holds\cite{We06}. In this paper, the definition of strong spatial
mixing is in the sense of Weitz.

Recently the  strong spatial mixing is also studied through
recursive formula. This approach is introduced by Weitz\cite{We06}
and Bandyopadhyay, Gamarnik \cite{BG06} for counting the number of
independent sets and colorings. The key point of this method is to
build the strong spatial mixing  on certain rooted trees. In
\cite{We06}, the equivalence between the marginal probability of a
vertex in a general graph $G$ and that of the root of a tree for
hard core model is proved using the $self$-$avoiding$ $tree$
technique. This shows the correlations on any graph decay at least
as fast as its corresponding self-avoiding tree. The strong spatial
mixing for hard-core model on bounded degree trees is also proved.
Later Gamarnik et.al.\cite{GK07} and Bayati et.al.\cite{BGKNT06}
bypass the construction of a self-avoiding tree. Instead, they
create a  $computation$ $tree$ and establish the strong spatial
mixing on the corresponding computation tree for list coloring and
matching problems.  Considering the Weitz's motivation of
construction of the self-avoiding tree, Jung and Shah\cite{JS06} and
Nair and Tetali \cite{NT07} generalize Weitz's work to certain
Markov random fields models, and Lu et.al.\cite{LMM07} on TP
decoding problem. Very recently Mossel and Sly\cite{MS08} show that
ferromagnetic Ising model exhibits strong correlation decay on
`sparse on average' graph under the tight assumption.

We consider the Binary Markov random fields, which are  also known
as two state spin systems, on a sparse on average graph where the
total degrees along each self-avoiding path (a path with distinct
vertices) with length $O(\log n)$ is $O(\log n)$ \cite{MS08}.  We
prove, for any `inverse temperature' on this graph, Gibbs
distribution exhibits strong spatial mixing when the `external
field' is uniformly larger than $B(d,\alpha_{\max},\gamma)$ or
smaller than $-B(d,-\alpha_{\min},\gamma)$. Here, $d$ is `maximum
average degree' and $\alpha_{\min}$, $\alpha_{\max}$, $\gamma$ are
parameters of the system. To the best of our knowledge, this
condition on `external field' is first considered for strong spatial
mixing. Our proof is based on a well known recursive formula
\cite{JS06} on a tree and the self-avoiding tree technique. We also
employed Lipchitz method, which was used in
\cite{BG06,BGKNT06,GK07}. The novelty of our proof is that we
propose a `path' characterization of Lipchitz method, which enables
us to give the `external field' condition in terms of `maximum
average degree' for the strong spatial mixing.

The remainder of the paper has the following structure. In Section
2, we present some preliminary definitions and notations. We go on
to propose the main result in Section 3. Section 4 is devoted to
prove the main theorem. Conclusion and further work  are given in
Section 5.\\\\
\textbf{\large{2. Preliminaries}}

Let $G=(V,E)$ be a finite graph with vertices $V=\{1,2,\cdots,n\}$
and edge set $E$, and let $d(u,v)$ denote the distance between $u$
and $v$, for any $u$, $v\in V$. A path $v_1,v_2,\cdots$ is called a
self-avoiding path if $v_i\neq v_j$ for any $i\neq j$. The distance
between a vertex $v\in V$ and a subset $\Lambda\subset V$ is defined
as
$$d(v,\Lambda)=\min\{d(v,u):u\in\Lambda\}.$$ A set of vertices with
distance  $l$ to the vertex $v$ is denoted by $$S(G,v,l)=\{u:
d(v,u)= l\}.$$
Let $\delta_v$ denote the degree of the vertex $v \in G$. The
$maximal$ $path$ $density$  of the graph $G$ is given by
$$m=m(G,v,l)=\max_{\Gamma}\sum_{u\in \Gamma}\delta_u,$$ where the
maximum is taken over all self-avoiding paths $\Gamma$ starting at
$v$ with length at most $l$. The $maximum$ $average$ $path$ $degree$
$\delta(G,v,l)$ is defined by $$\delta(G,v,l)=(m(G,v,l)-\delta_v)/l,
\qquad l\geq 1. $$ The $maximum$ $average$ $degree$ of $G$ is
defined as $$\Delta(G,l)=\max_{v\in V}\delta(G,v,l).$$ For any order
of all the vertices in $G$ given, an  associated partial order of
$E$ based on  the order of $V$ defined as $(i,j)>(k,l)$ if and only
if $(i,j)$ and $(k,l)$ share a common vertex and $i+j > k+l$. In
binary Markov random fields(BMRF) on $G$, each vertex $i\in V$ is
associated with a random variable $X_i$ with range
$\Omega=\{{\pm1}\}$(briefly $\pm$).\\\\
{\textbf{Definition 1.}} The Gibbs measure of BMRF on $G$ is defined
by the joint distribution of the random variable
$X=\{X_1,X_2,\cdots,X_{n}\}$
\begin{displaymath} P_G(X=\sigma)=\frac{1}{Z(G)}\exp(\sum\limits_{(i,j)\in
E}\beta_{ij}(\sigma_i,\sigma_j)+\sum\limits_{i\in V}h_i(\sigma_i)),
\end{displaymath}
where $h_i:\Omega\rightarrow R$ and $\beta_{ij}:\Omega^2\rightarrow
R$. Here $Z(G)$ is called the  partition function of the system.

\

Note that the Gibbs measure would satisfy  
$\sum_{\sigma\in\Omega^n}P_G(X=\sigma)=1$. We use notation
$\beta_{ij}(a,b)=\beta_{ji}(b,a)$. For any $\Lambda\subseteq V$,
$\sigma_{\Lambda}$ denotes the set $\{ \sigma_i, i\in \Lambda\}$.
With a little abuse of notation, $\sigma_{\Lambda}$ also denotes the
condition or configuration that  $i$ is fixed $\sigma_i$, for any $
i\in\Lambda$. Let $Z(G,\Phi)$ denote the partition function under
the condition $\Phi$, e.g. $Z(G,X_1=+)$ represent the partition
function under the condition the vertex $1$ is fixed $+$.\\
\begin{figure}[ht]
\centering
\includegraphics[scale=.4]{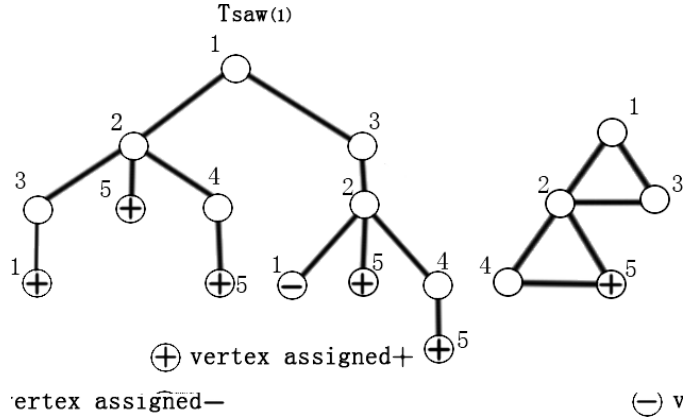}
\caption{\small  The graph with one vertex assigned + (Right)  and
its corresponding self-avoiding tree $T_{saw(1)}$ (Left)}
\end{figure}

A self-avoiding walk (SAW) is a sequence of moves (on a graph) which
does not visit the same point more than once. The following gives an
important tool in proving our results. It is introduced in
\cite{We06}.

\

\noindent \textbf{Definition 2.} (Self-Avoiding Tree) The
self-avoiding tree $T_{saw(v)}(G)$ (for simplicity denoted by
$T_{saw(v)}$) corresponding to the vertex $v$ of $G$ is the tree
with root $v$ and generated through the self-avoiding walks
originating at $v$. A vertex closing a cycle is included as a leaf
of the tree and is assigned to be $+$, if the edge ending the cycle
is larger
than the edge starting the cycle, and $-$ otherwise.\\

\noindent \emph{Remark:}  Given any configuration $\sigma_{\Lambda}$
of $G$, $\Lambda\subset V$, the self-avoiding tree is constructed
the same as the above procedure except that,  the vertex  which is a
copy of the vertex $i$ in $\Lambda$ is fixed to the same spin
$\sigma_i$ as $i$ and the subtree below it is not constructed due to
the Markov property, see Figure 1 for example, where vertex $5$ is
fixed $+$ in $G$. 
\\

To generalize the strong spatial mixing property on trees to general
graph,  we  need to utilize the remarkable property of the
self-avoiding tree, one of two main results of \cite{We06}, and
explicitly stated in \cite{JS06}. For any configuration
$\sigma_{\Lambda}$ of $G$, $\Lambda\subset V$, we also use
$\sigma_{\Lambda}$ to denote the configuration of $T_{saw(v)}$
obtained by imposing the condition corresponding to
$\sigma_{\Lambda}$.\\\\
\textbf{Proposition 1.}\emph{ For BMRF on $G=(V,E)$, for any
configuration $\sigma_{\Lambda}$ on $G$, $\Lambda\subset V$ and any
vertex $v\in V$, then
\begin{displaymath}
P_G(X_v=+|\sigma_{\Lambda})=P_{T_{saw(v)}}(X_v=+|\sigma_{\Lambda}).
\end{displaymath}}\\

In order to study results to the sparse on average graph, their
following properties are useful. The proof is based on induction
and can be found in \cite{MS08}.\\\\
\textbf{Proposition 2.} \emph{ Let $j$, $l$ be positive integers.
Then one has
\begin{displaymath}m(G,v,jl)\leq j\max\limits_{u\in
G}\{m(G,u,l)-\delta_u\}+\delta_v \end{displaymath}
and 
\begin{displaymath}
|S(T_{saw(v)},v,l+1)|\leq\delta_v(\delta(G,v,l)-1)^l.
\end{displaymath}}\\\\
\textbf{Definition 3.} (Strong Spatial Mixing) The Gibbs
distribution of BMRF exhibits strong spatial mixing if and only if
there exist positive numbers  $a$, $b$, $c$ independent of $n$, for
any vertex $v\in V$, subset $\Lambda\subset V$, any two
configurations $\sigma_{\Lambda}$ and $\eta_{\Lambda}$ on $\Lambda$,
denote perturbation set $\Theta=\{v\in\Lambda: \sigma_v\neq
\eta_v\}$ and $t=d(v,\Theta)$, when $t=ka\log n+1$, $k=1,2,\cdots$,
\begin{displaymath}
|P_G(X_v=+|\sigma_{\Lambda})-P_G(X_v=+|\eta_{\Lambda})|\leq
f(t),\end{displaymath} where decay function $f(t)= b\exp(-c t)$.\\\\
\textbf{\large{3. Main Results}}



In the binary Markov random fields, it is well known that if
$\beta_{ij}(\sigma_i,\sigma_j)=J_{ij}\sigma_i\sigma_j$ and
$h_i=B_i\sigma_i$ for all the edge $(i,j)\in E$ and vertex $i\in V$,
and $J_{ij}$ is uniformly  positive (or negative) for all $(i,j)\in
E$, the BMRF is called ferromagnetic (or antiferromagnetic) Ising
model. For simplicity, we use the following notations. Let
$$J_{ij}=\frac{\beta_{ij}(+,+)+\beta_{ij}(-,-)-\beta_{ij}(-,+)-\beta_{ij}(+,-)}{4},$$
and $B_i=\frac{h_i(+)-h_i(-)}{2}$ for all edges and vertices . We
call $J_{ij}$ and $B_i$ `inverse temperature' and `external field'
of BMRF. Let $J=\max_{(i,j)\in E}|J_{ij}|$, $B_{\min}=\min_{i\in
V}B_{i}$, and $B_{\max}=\max_{i\in V}B_{i}$. Denote
$$\alpha_{\max}=\max\limits_{(i,j)\in
E}\{\beta_{ij}(-,-)-\beta_{ij}(+,-),\beta_{ij}(-,+)-\beta_{ij}(+,+)\}$$
and $$\alpha_{\min}=\min\limits_{(i,j)\in
E}\{\beta_{ij}(-,-)-\beta_{ij}(+,-),\beta_{ij}(-,+)-\beta_{ij}(+,+)\}.$$
Let $$\gamma_{ij}=\max_{(i,j)\in
E}\{\frac{|b_{ij}c_{ij}-a_{ij}d_{ij}|}{a_{ij}c_{ij}},
\frac{|b_{ij}c_{ij}-a_{ij}d_{ij}|}{b_{ij}d_{ij}}\},$$ and
$\gamma=\max_{(i,j)\in E}\{\gamma_{ij}\}$, where

\qquad $a_{ij}=\exp(\beta_{ij}(+,+))$, \qquad
$b_{ij}=\exp(\beta_{ij}(+,-))$,

\qquad $c_{ij}=\exp(\beta_{ij}(-,+))$, \qquad
$d_{ij}=\exp(\beta_{ij}(-,-))$. \\\\
\textbf{Theorem 1.} \emph{ Let $G=(V,E)$ be a graph with $n$
vertices. 
There exit two positive numbers $a>0$ and $d>0$ such that
$\Delta(G,a\log n)\leq d$, and $(d-1)\tanh{J}\geq1$. Assume
\begin{displaymath}
B_{\min}>B(d,\alpha_{\max},\gamma)\ \ \ \ \ \ or \ \ \ \ \
B_{\max}<-B(d,-\alpha_{\min},\gamma)
\end{displaymath}
where
$$B(d,\alpha,\gamma)=\frac{(d-1)\alpha}{2}+\log(\frac{\sqrt{\gamma(d-1)}+\sqrt{\gamma(d-1)-4}}{2}).$$
Then the Gibbs distribution of BMRF exhibits exponential strong
spatial
mixing.}\\

\noindent \emph{Remark:} In Theorem 1, by the definition of
$\gamma_{ij}$, one has $$\gamma_{ij}^2\geq
|(\frac{b_{ij}}{a_{ij}}-\frac{d_{ij}}{c_{ij}})(\frac{c_{ij}}{d_{ij}}-\frac{a_{ij}}{b_{ij}})|=(e^{2J_{ij}}-e^{-2J_{ij}})^2,$$
hence,
$$\gamma_{ij}\geq|e^{2J_{ij}}-e^{-2J_{ij}}|=(e^{J_{ij}}+e^{-J_{ij}})^2\tanh
J_{ij}\geq 4\tanh J_{i,j}.$$ Therefore, $\gamma(d-1)-4\geq 0$ under
the condition $(d-1)\tanh{J}\geq1$ . The case of $(d-1)\tanh{J}<1$
is discussed separately in \cite{ZLB09} with totally different
method. The decay function corresponding to the above two conditions
are respectively
\begin{displaymath}
f(t)=\frac{\delta_i\gamma}{4}(\frac{(d-1)\gamma\exp(2B_{\min}-(d-1)\alpha_{\max})}{(1+\exp(2B_{\min}-(d-1)\alpha_{\max}))^2})^{t-1}
\end{displaymath}
and 
\begin{displaymath}
f(t)=\frac{\delta_i\gamma}{4}(\frac{(d-1)\gamma\exp(2B_{\max}-(d-1)\alpha_{\min})}{(1+\exp(2B_{\max}-(d-1)\alpha_{\min}))^2})^{t-1}.
\end{displaymath}
\\\\
\textbf{\large{4. Proofs}}

 Theorem 1 is proved with the recursive
formula\cite{JS06}. The technique used is Lipchitz method, which is
well known. A `path' version of it is presented first. We use the
following notations for simplicity. Let $T=(V,E)$ be a tree rooted
at $0$ with vertices $V=\{0,1,2,\cdots,n\}$, edge set $E$ and BMRF
on it. For each edge $(i,j)\in E$, recall the notation in Theorem 1,

$a_{i,j}=e^{\beta_{ij}(+,+)}$, $b_{i,j}=e^{\beta_{ij}(+,-)}$,
$c_{i,j}=e^{\beta_{ij}(-,+)}$, and $d_{i,j}=e^{\beta_{ij}(-,-)}$.

\noindent Let $M_{ij}=c_{ij}-d_{ij}$ and $N_{ij}=a_{i,j}-b_{i,j}$.
Define
\begin{displaymath}
f_{ij}(x)=\frac{M_{ij}x+d_{ij}}{N_{ij}x+b_{ij}}\ \ \ \
\operatorname{and}\ \ \ \
h_{ij}(x)=\frac{a_{ij}d_{ij}-b_{ij}c_{ij}}{(M_{ij}x+d_{ij})(N_{ij}x+b_{ij})}.
\end{displaymath} For any $i\in V$, let
$T_i$ denote the subtree rooted at $i$ and there is a associated
BMRF on $T_i$ restricted by BMRF on $T$. Recall
$B_{i}=\frac{h_{i}(+)-h_i(-)}{2}$ is the external field. Denote
$\lambda_{i}=e^{-2B_i}$, and let $\Gamma_{ij}$ be the unique
self-avoiding path from $i$ to $j$ on $T$.\\\\
\textbf{Lemma 1.} \emph{For any $(i,j)\in E$,
$\max\limits_{x\in[0,1]}|h_{ij}(x)|\leq \gamma_{ij}$.
}\\\\
\textbf{Proof.} Since $M_{ij}x+d_{ij}\geq 0$ and $N_{ij}x+b_{ij}\geq
0$, $\forall x\in[0,1]$, thus all we need is to show
$$\min\limits_{x\in[0,1]}w(x)=\min(a_{ij}c_{ij}, b_{ij}d_{ij}), $$
where $w(x)=(M_{ij}x+d_{ij})(N_{ij}x+b_{ij})$. The case
$M_{ij}N_{ij}=0$ is trivial. Hence without loss of generality,
suppose $M_{ij}N_{ij}\neq0$. Noting
$x_l=-\frac{d_{ij}N_{ij}+b_{ij}M_{ij}}{2M_{ij}N_{ij}}$ is an
extremum of $w(x)$ on $R$. There are three cases needed to be discussed.\\
Case 1. $M_{ij}N_{ij}<0$, then $w(x)$ reaches its minimum at
boundary. Then
$\min\limits_{x\in[0,1]}w(x)\leq\min(w(0),w(1))=\min(a_{ij}c_{ij},
b_{ij}d_{ij})$.\\
Case 2. $M_{ij}>0, N_{ij}>0$, then $x_l\leq 0$, $w(x)$ is increasing
on $[0,1]$, then $\min\limits_{x\in[0,1]}w(x)=w(0)=b_{ij}d_{ij}$.\\
Case 3. $M_{ij}<0, N_{ij}<0$, then $x_l\geq 1$, $w(x)$ is decreasing
on $[0,1]$, hence $\min\limits_{x\in[0,1]}w(x)=w(1)=a_{ij}c_{ij}$. \
\ \ \ $\Box$\\

With Lemma 1, we present a `path' version of Lipchitz
approach.\\\\
\textbf{Lemma 2.}\emph{ Let $\Lambda\subset V$ , $\zeta_{\Lambda}$
and $\eta_{\Lambda}$ be any two configurations on $\Lambda$. Let
$\Theta=\{i: \zeta_i\neq \eta_i, i\in\Lambda\}$, $t=d(0,\Theta)$ and
$S(T,0,t)=\{i:
 d(0,i)=t, i\in T\}$. Then
\begin{displaymath}
\begin{split}
&|P_T(X_0=+|\zeta_{\Lambda})-P_T(X_0=+|\eta_{\Lambda})|\\
&\leq \gamma^{t}\sum\limits_{k\in
S(T,0,t)}\prod\limits_{\substack{i\in \Gamma_{0k}\\ i\neq
k}}g_{i}(z_i)(1-g_{i}(z_i))
\end{split}
\end{displaymath}
where $z_i$ are constant vectors with elements in $[0,1]$, and
$g_{i}(x_i)=(1+\lambda_i\prod\limits_{(i,i_j)\in
T_i}f_{ii_j}(x_{ii_j}))^{-1}$,
$x_i=(x_{ii_1},x_{ii_2},\cdots,x_{ii_{\delta_i-1}})$. }\\\\
\textbf{Proof.} For any vertex $i$ in $T$, recall  $T_i$ denote the
subtree rooted at $i$  with BMRF induced on $T_i$ by $T$. Let
$p^{\zeta_{\Lambda}}_{i}\equiv P_{T_i}(X_i=+|\zeta_{\Lambda_i})$ and
$R^{\zeta_{\Lambda}}_i\equiv
\frac{P_{T_i}(X_i=+|\zeta_{\Lambda_i})}{P_{T_i}(X_i=-|\zeta_{\Lambda_i})}$,
where $\zeta_{\Lambda_i}$ is configuration by restriction of
$\zeta_{\Lambda}$ on $T_i$. Let $\Omega_{T_i}$ denote the
configuration space in $T_i$ under the condition $\zeta_{\Lambda}$,
$i=1,2,\cdots,n$. $\Omega_{0}$ denotes the configuration space of
$T_0$ under the condition $\zeta_{\Lambda}\cup\{\sigma_0\}$. Let
$0_1,0_2,\cdots,0_q$ be the neighbors connected to $0$,
$q=\delta_0$(the degree of the root). Now we present the recursive
formula,
\begin{displaymath}
\begin{split}
&R^{\zeta_{\Lambda}}_0=\frac{Z(T_0,X_0=+,\zeta_{\Lambda})}{Z(T_0,X_0=-,\zeta_{\Lambda})}\\
&=\frac{e^{h_0(+)}\sum\limits_{\sigma\in\Omega_0}e^{\sum\limits_{i=1}^q(\beta_{00_i}(+,\sigma_{0_i})+\sum\limits_{(k,l)\in
T_{0_i}}\beta_{kl}(\sigma_k,\sigma_l)+\sum\limits_{k\in
T_{0_i}}h_k(\sigma_k))}}{e^{h_0(-)}\sum\limits_{\sigma\in\Omega_0}e^{\sum\limits_{i=1}^q(\beta_{00_i}(-,\sigma_{0_i})+\sum\limits_{(k,l)\in
T_{0_i}}\beta_{kl}(\sigma_k,\sigma_l)+\sum\limits_{k\in T_{0_i}}h_k(\sigma_k))}}\\
&=e^{2B_0}\prod\limits^{q}_{i=1}\frac{\sum\limits_{\sigma\in\Omega_{T_{0_i}}}e^{\beta_{00_i}(+,\sigma_{0_i})+\sum\limits_{(k,l)\in
T_{0_i}}\beta_{kl}(\sigma_k,\sigma_l)+\sum\limits_{k\in
T_{0_i}}h_k(\sigma_k)}}{\sum\limits_{\sigma\in\Omega_{T_{0_i}}}e^{\beta_{00_i}(-,\sigma_{0_i})+\sum\limits_{(k,l)\in
T_{0_i}}\beta_{kl}(\sigma_k,\sigma_l)+\sum\limits_{k\in
T_{0_i}}h_k(\sigma_k)}}\\
\end{split}
\end{displaymath}
\begin{displaymath}
\begin{split}
&=e^{2B_0}
\prod\limits^{q}_{i=1}\frac{a_{00_i}Z(T_{0_i},X_i=+,\zeta_{\Lambda_i})+b_{00_i}Z(T_{0_i},X_i=-,\zeta_{\Lambda_i})}
{c_{00_i}Z(T_{0_i},X_i=+,\zeta_{\Lambda_i})+d_{00_i}Z(T_{0_i},X_i=-,\zeta_{\Lambda_i})}\\
&=e^{2B_0} \prod\limits^{q}_{i=1}\frac{a_{00_i}
R^{\zeta_{\Lambda}}_{0_i}+b_{00_i}}{c_{00_i}
R^{\zeta_{\Lambda}}_{0_i}+d_{00_i}}.
\end{split}
\end{displaymath}
Then we have the following equality
\begin{displaymath}
\begin{split}
p^{\zeta_{\Lambda}}_{0}&=P_T(X_0=+|\zeta_{\Lambda})\\
&=\frac{1}{1+\frac{P_T(X_0=-|\zeta_{\Lambda})}{P_T(X_0=+|\zeta_{\Lambda})}}
=\frac{1}{1+1/R^{\zeta_{\Lambda}}_0}\\
&=\frac{1}{1+\lambda_0\prod\limits_{(0,0_j)\in
T}\frac{c_{00_j}R^{\zeta_{\Lambda}}_{0_j}+d_{00_j}}{a_{00_j}
R^{\zeta_{\Lambda}}_{0_j}+b_{00_j}}}\\
&=\frac{1}{1+\lambda_0\prod\limits_{(0,0_j)\in
T}\frac{M_{00_j}p^{\zeta_{\Lambda}}_{0_j}+d_{00_j}}{N_{00_j}p^{\zeta_{\Lambda}}_{0_j}+b_{00_j}}}\\
&=g_0(x_0),
\end{split}
\end{displaymath}
where $x_0=(p^{\zeta_{\Lambda}}_{0_1},
p^{\zeta_{\Lambda}}_{0_2},\cdots,p^{\zeta_{\Lambda}}_{0_{\delta_0}})$.
First, note that for any $x=(x_1,x_2,\cdots,x_q)$ and
$y=(y_1,y_2,\cdots,y_q)$, first order Taylor expansion at $y$ gives
that there exists a $\theta\in[0,1]$ such that
\begin{displaymath}
g_0(x)-g_0(y)=\nabla g_0(y+\theta (x-y))(x-y)^{T},
\end{displaymath}
where $(x-y)^{T}$ denotes the transportation of the vector $(x-y)$.
Careful calculations give the following \\
\begin{displaymath}
\begin{split}
\frac{\partial g_0(x)}{\partial
x_i}&=-\frac{\lambda_0\prod\limits^q_{j=1}f_{00_j}(x_{j})
(\frac{d\log(f_{00_i}(x_{i})}{dx_i})}
{(1+\lambda_0\prod\limits^q_{j=1}f_{00_j}(x_{j}))^{2}}\\
&=-g_0(x)(1-g_0(x))(\frac{M_{00_i}}{M_{00_i}x_i+d_{00_i}}-\frac{N_{00_i}}{N_{00_i}x_i+b_{00_i}})\\
&=g_0(x)(1-g_0(x))\frac{a_{00_i}d_{00_i}-b_{00_i}c_{00_i}}{(M_{00_i}x_i+d_{00_i})(N_{00_i}x_i+b_{00_i})}\\
&=g_0(x)(1-g_0(x))h_{00_i}(x_i).
\end{split}
\end{displaymath}
Hence, let $x_0=(p^{\zeta_{\Lambda}}_{0_1},
p^{\zeta_{\Lambda}}_{0_2},\cdots,p^{\zeta_{\Lambda}}_{0_{\delta_0}})$
and $y_0=(p^{\eta_{\Lambda}}_{0_1},
p^{\eta_{\Lambda}}_{0_2},\cdots,p^{\eta_{\Lambda}}_{0_{\delta_0}})$,
then there exits $\theta_0\in [0,1]$ such that
\begin{equation}
\begin{split}
|p^{\zeta_{\Lambda}}_{0}-p^{\eta_{\Lambda}}_{0}|&\leq
\sum\limits_{j=1}^{q}|g_0(z_0)(1-g_0(z_0))h_{00_j}(x_j)||p^{\zeta_{\Lambda}}_{0_j}-p^{\eta_{\Lambda}}_{0_j}|\\
&\leq
\sum\limits_{j=1}^{q}g_0(z_0)(1-g_0(z_0))\gamma_{00_j}|p^{\zeta_{\Lambda}}_{0_j}-p^{\eta_{\Lambda}}_{0_j}|\\
&\leq \gamma
\sum\limits_{j=1}^{q}g_0(z_0)(1-g_0(z_0))|p^{\zeta_{\Lambda}}_{0_j}-p^{\eta_{\Lambda}}_{0_j}|,
\end{split}
\end{equation}
where $z_0=x_{0}+\theta_0(x_{0}-y_{0})$ and the second inequality
follows by Lemma 1. Now repeat the procedure on the subtree
$T_{0_j}$ for
$|p^{\zeta_{\Lambda}}_{0_j}-p^{\eta_{\Lambda}}_{0_j}|$,
$j=1,2,\cdots,q$ and so on.  We can see that the summation is over
all the self-avoiding paths starting at the root $0$. For each path
$\Gamma$, if the end point of $\Gamma$ is a leave $j$ with
$d(0,j)\leq t-1$ or there is a vertex $i$ on $\Gamma$ with
$d(0,i)\leq t-1$ being fixed, the contribution of the path to the
summation is zero since
$p^{\zeta_{\Lambda}}_{i}-p^{\eta_{\Lambda}}_{i}=p^{\zeta_{\Lambda}}_{j}-p^{\eta_{\Lambda}}_{j}=0$.
Hence the remaining path with length $t$ is  in the set
$\{\Gamma_{0k}: k\in S(T,0,t) \}$. This completes the proof of lemma
2. \ \ \ \ $\Box$\\

In order to complete the proof of Theorem 1, we need the following lemma.\\\\
\textbf{Lemma 3.} \emph{ Let $\lambda_i\geq0$, $i=1,2,\cdots,n$.
Then
\begin{displaymath}
\prod\limits_{i=1}^n(1+\lambda_i)\geq
(1+\sqrt[n]{\prod\limits^n_{i=1}\lambda_i})^n.
\end{displaymath}}\\\\
\textbf{Proof.} Consider  \begin{displaymath}
\begin{split}
\prod\limits_{i=1}^n(1+\lambda_i)&=1+\sum\limits^n_{k=1}
(\sum\limits_{i_1<i_2<\cdots<i_k}\prod\limits^k_{j=1}\lambda_{i_j})\\
&\geq 1+\sum\limits^n_{k=1}(C^k_n(\prod\limits^n_{i=1}\lambda_i)^{\frac{C_{n-1}^{k-1}}{C^k_n}})\\
&=1+\sum\limits^n_{k=1}(C^k_n(\prod\limits^n_{i=1}\lambda_i)^{\frac{k}{n}})\\
&=(1+\sqrt[n]{\prod\limits^n_{i=1}\lambda_i})^n,
\end{split}
\end{displaymath}
where $C^k_n=\frac{n!}{k!(n-k)!}$. The first inequality uses the
arithmetic-geometric average inequality.\\

With Lemma 2 and 3, it is sufficient to prove Theorem 1.\\\\
\textbf{Proof of Theorem 1.} Following the notation of Lemma 2, let
$s=|S(T,0,t)|$,
we have \\
\begin{displaymath}
\begin{split}
|p^{\zeta_{\Lambda}}_{0}-p^{\eta_{\Lambda}}_{0}|&\leq
\gamma^{t}\sum\limits_{k\in S(T,0,t)}\prod\limits_{\substack{i\in
\Gamma_{0k}\\
i\neq k}}g_{i}(z_i)(1-g_{i}(z_i))\\
&\leq s\gamma^{t} \max\limits_{\substack{k\in
S(T,0,t)}}\prod\limits_{\substack{i\in \Gamma_{0k}\\ i\neq
k}}g_{i}(z_i)(1-g_{i}(z_i))\\
&\leq s\frac{\gamma^{t}}{4}\max\limits_{\substack{(0,0_j)\in T\\k\in
S(T,0,t)}}\prod\limits_{\substack{i\in \Gamma_{0_jk}\\ i\neq
k}}g_{i}(z_i)(1-g_{i}(z_i)).
\end{split}
\end{displaymath}
For each $\Gamma_{0_jk}$, where $(0,0_j)\in T$, $k\in S(T,0,t)$,
\begin{displaymath}
\begin{split}
&\prod\limits_{\substack{i\in \Gamma_{0_jk}\\ i\neq
k}}g_{i}(z_i)(1-g_{i}(z_i))\\&=\prod\limits_{\substack{i\in
\Gamma_{0_jk}\\ i\neq k}}\frac{\lambda_i\prod\limits_{(i,i_l)\in
T_i}f_{ii_l}(z_{ii_l})}
{(1+\lambda_i\prod\limits_{(i,i_l)\in T_i}f_{ii_l}(z_{ii_l}))^{2}}\\
&\leq (\frac{r_{jk}} {(1+r_{jk})^2})^{t-1},
\end{split}
\end{displaymath}
where $r_{jk}=(\prod\limits_{i\in \Gamma_{0_jk} i\neq
k}\lambda_i\prod\limits_{(i,i_l)\in
T_i}f_{ii_l}(z_{ii_l}))^{1/(t-1)}$ and the inequality above follows
from Lemma 3.  A simple calculation gives that
$e^{\alpha_{\min}}\leq f_{ij}(x)\leq e^{\alpha_{\max}}$, for any
$(i,j)\in T$. Hence,
\begin{displaymath}
\begin{split}
e^{\alpha_{\min}(\delta(T,0,t-1)-1)}&\leq (
\prod\limits_{\substack{i\in \Gamma_{0_jk}\\ i\neq
k}}\prod\limits_{(i,i_l)\in T_i}f_{ii_l}(z_{ii_l}))^{1/(t-1)}\leq
e^{\alpha_{\max}(\delta(T,0,t-1)-1)}.
\end{split}
\end{displaymath}
Now we prove the exponential strong spatial mixing under assumption
of Theorem 1. Suppose $\Gamma$ is a self-avoiding path of $G$.
Noting that each self-avoiding path on $T$ by removing the ending
point is also a self-avoiding path on $G$. From proposition 1, we
know $0$ is a vertex of $G$ and let
$\bar{p}_0^{\zeta_{\Lambda}}=P_G(X_0=+|\zeta_{\Lambda})$. By
proposition 2, we know $\delta(T,0,t-1)\leq \Delta(G,t-1)\leq d$
when $t=ka\log n+1$, $k=1,2,\cdots$. If $B_{\min}>
B(d,\alpha_{\max},\gamma)$, then
\begin{displaymath}
 \frac{\gamma(d-1)\exp(2B_{\min}-\alpha_{\max}(d-1))}
{(1+\exp(2B_{\min}-\alpha_{\max}(d-1)))^2}<1.
\end{displaymath}
By proposition 2, we know $s\leq \delta_0(d-1)^{t-1}$. Noting
$(\prod\limits_{\substack{i\in \Gamma_{0_jk}\\ i\neq
k}}\lambda_i)^{1/(t-1)}\leq e^{-2B_{\min}}$ , now we can see
\begin{displaymath}
\begin{split}
&|\bar{p}_0^{\zeta_{\Lambda}}-\bar{p}_0^{\eta_{\Lambda}}|=|p^{\zeta_{\Lambda}}_{0}-p^{\eta_{\Lambda}}_{0}|\leq
s\frac{\gamma^t}{4}(\frac{r_{jk}} {(1+r_{jk})^2})^{t-1}\\
&\leq
\frac{\delta_{0}\gamma}{4}(\frac{\gamma(d-1)\exp(2B_{\min}-\alpha_{\max}(d-1))}
{(1+\exp(2B_{\min}-\alpha_{\max}(d-1)))^2})^{t-1},
\end{split}
\end{displaymath}
where the first equality follows from the proposition 1. The similar
case holds for $B_{\max}< -B(d,-\alpha_{\min},\gamma)$.
This completes the proof.   \ \ \ \  $\Box$\\

From the proof above, we can see if the graph is bounded degree with
maximum degree is $d$, the condition for `external field' can be
relaxed to $B_i>B(d,\alpha_{\max},\gamma)$ or $
B_i<-B(d,-\alpha_{\min},\gamma)$ for any $i\in V$, which does not
require that `external field' is uniformly large or uniformly
small as in Theorem 1. \\\\\
\textbf{Corollary 1.} \emph{Let $G=(V,E)$ be a bounded graph and
with maximum degree $d$ and BMRF on it, and $\tanh J(d-1)\geq1$. If
$B_i>B(d,\alpha_{\max},\gamma)$ or $B_i<-B(d,-\alpha_{\min},\gamma)$
for any $i\in V$. Then the Gibbs distribution exhibits strong
spatial
mixing.}\\\\
\textbf{Proof.} Following the notations above, by the formula (1) in
Lemma 2, we have
\begin{displaymath}
\begin{split}
|p^{\zeta_{\Lambda}}_{0}-p^{\eta_{\Lambda}}_{0}|&\leq \gamma
\sum\limits_{j=1}^{q}g_0(z_0)(1-g_0(z_0))|p^{\zeta_{\Lambda}}_{0_j}-p^{\eta_{\Lambda}}_{0_j}|.\\
\end{split}
\end{displaymath}
Without loss of generality, suppose the degree of $0$ is $d-1$. Then
\begin{displaymath}
\begin{split}
|p^{\zeta_{\Lambda}}_{0}-p^{\eta_{\Lambda}}_{0}|&\leq \gamma(d-1)
\max_{i\in T}(g_0(z_i)(1-g_0(z_i)))|p^{\zeta_{\Lambda}}_{0_j}-p^{\eta_{\Lambda}}_{0_j}|.\\
\end{split}
\end{displaymath}
If $B_i>B(d,\alpha_{\max},\gamma)$ or
$B_i<-B(d,-\alpha_{\min},\gamma)$, we know
$\gamma(d-1)g_{i}(z_i)(1-g_{i}(z_i))<1$ for any $i\in T$. Hence by
induction on the hight $t$, we get
\begin{displaymath}
|p^{\zeta_{\Lambda}}_{0}-p^{\eta_{\Lambda}}_{0}|\leq (\gamma(d-1)
\max_{i\in T}(g_0(z_i)(1-g_0(z_i))))^{t}.
\end{displaymath}
Since the degree of $0$ is at most $d$. Then
\begin{displaymath}
\begin{split}
|p^{\zeta_{\Lambda}}_{0}-p^{\eta_{\Lambda}}_{0}|&\leq  \gamma d
g_0(z_0)(1-g_0(z_0))(\gamma(d-1) \max_{i\in
T}(g_0(z_i)(1-g_0(z_i))))^{t-1}\\&\leq \frac{\gamma
d}{4}(\gamma(d-1) \max_{i\in T}(g_0(z_i)(1-g_0(z_i))))^{t-1}.
\end{split}
\end{displaymath}
\begin{figure}[ht]
\centering
\includegraphics[scale=.6]{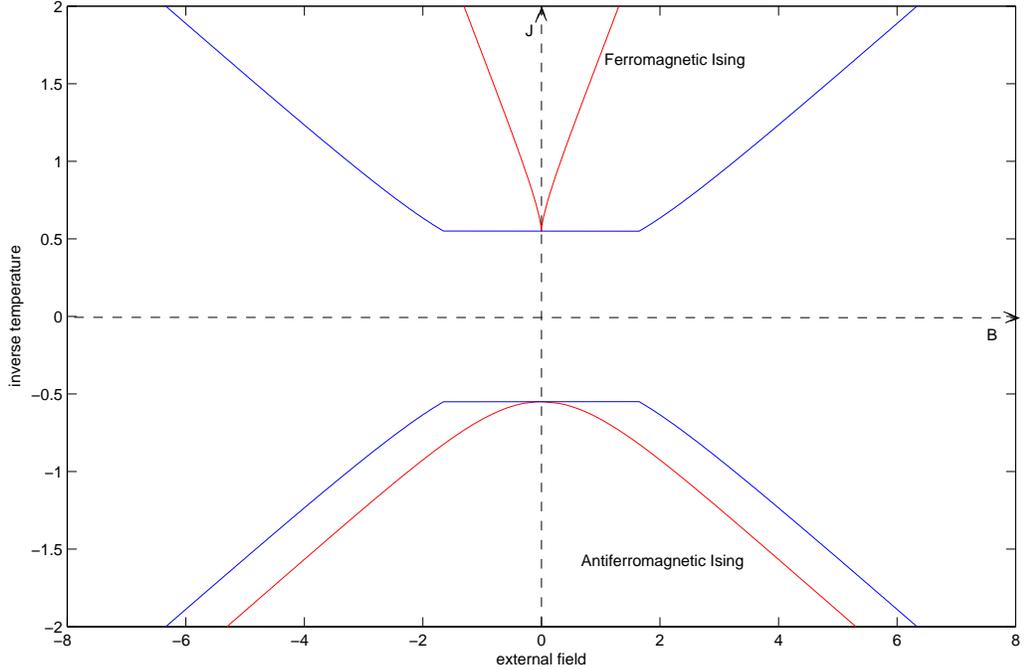}
\caption{\small{The red line and the blue line denote the ``external
field" curves for uniqueness of Gibbs measures and  for strong
spatial mixing on three regular trees $T$ respectively, where
$\beta_{ij}(\sigma_{i},\sigma_{j})=J\sigma_{i}\sigma_{j}$ and
$h_i(\sigma_{i})=B\sigma_{i}$}, for any $i\in T$ and $(i,j)\in T$.}
\end{figure}
Applying proposition 1 completes the proof.\ \ \ \ \ $\Box$\\

\emph{Remark:} We emphasize that the tighter bound of $f_{ij}(x)$ is
the key to improve the result since better bound of $f_{ij}(x)$ will
give better bound for $g_i(x)$. We do not optimize the parameter
here. We are not aware that Lipchitz method  can make
$B(d,\alpha_{\max},\gamma)$ or $-B(d,-\alpha_{\min},\gamma)$
optimally approximate the critical point of `external field' for
uniqueness of Gibbs measures if they does exit. Note that the
critical points of `external field' for ferromagnetic and
antiferromagnetic Ising model are different on  an infinite $d$
regular tree with  degree $d$ for each vertex \cite{Ge88}). We do
not expect the critical external field for Ising model on $d$
regular tree for uniqueness of Gibbs measures is the optimal
external field for strong spatial mixing. The intuition for this is
that the uniqueness of Gibbs measures on the tree is equivalent to
weak spatial mixing (see \cite{BMP99}\cite{We06} for definitions) in
some sense\cite{Ly89}. If some configurations are close to the
root(note some configurations may be at the hight $2$ or $3$(see
Figure 1) when self-avoiding tree is constructed), the perturbation
of the boundary condition changes the Gibbs measures at the root
radically. More precisely, strong spatial mixing can be deducted to
the weak spatial mixing by removing the support of unmodified
boundary configuration and changing the external field of some
vertices (see Lemma 2 in \cite{ZLB09}). Hence, the critical external
field condition for weak spatial mixing does not hold for strong
spatial mixing. Figure 2 illustrates the curve of external field
under our condition for strong spatial mixing and the critical
external field for uniqueness of Gibbs measures on infinite $d$
regular tree, where $d=3$.\\\\
\textbf{\large{5. Conclusion and Further Work}}

The Gibbs distribution on a graph $G=(V,E)$ with `maximum average
degree' $d$ is considered in this paper. The (exponential) strong
spatial mixing is proved for such systems,  when the `external
field' $B_i$ is uniformly larger than $B(d,\alpha_{\max},\gamma)$ or
smaller than $-B(d,-\alpha_{\min},\gamma)$. Here
$B(d,\alpha,\gamma)$ is a function with parameter $d$, $\alpha$,
$\gamma$. It is not difficult to apply our results to
Erd$\ddot{o}$-R$\dot{e}$nyi random graph $G(n,d/n)$, where each edge
is chosen independently with probability $d/n$\cite{MS08}.

For future work, some improvements to the condition on `external
field' should be possible. We have emphasized the essential key
points in the remark of last section. However, we believe that it
requires other method other than Lipchitz method. The fixed point
method in\cite{Ke85} may be a possible approach.



\begin{thebibliography}{19}

\bibitem{BG06}
A. Bandyopadhyay and D. Gamarnik.: Counting without sampling: New
algorithms for enumeration problems using statistical physics,
Proceedings of 17th ACM-SIAM Symposium on Discrete Algorithms (SODA)
(2006).

\bibitem{BGKNT06}
M. Bayati, D. Gamarnik, D. Katz, C. Nair, and P. Tetali.: Simple
deterministic approximation algorithms for counting matchings,
Proceedings of the 39th annual ACM symposium on Theory of
computing(STOC) (2007).

\bibitem{BMP99}
J.Bertoin, F.Martinelli and Y.Peres.: Lectures on Probability Theory
and Statistics, Ecole d'Et\'{e} de Probabilit\'{e}s de Saint-Flour
XXVII-1997, 92-191, (1999).


\bibitem{Ge88}
H. O. Georgii.: Gibbs measures and phase transitions, volume 9 of de
Gruyter Studies in Mathematics. Walter de Gruyter $\&$ Co., Berlin,
(1988).

\bibitem{GK07}
D. Gamarnik and D. Katz.: Correlation decay and deterministic FPTAS
for counting list-colorings of a graph, Proceedings of 18th ACM-SIAM
Symposium on Discrete Algorithms (SODA) (2007).

\bibitem{JS06}
K. Jung and D. Shah.: Inference in Binary Pair-wise Markov Random
Field through Self-Avoiding Walk, Preprint on
http://arxiv.org/abs/cs.AI/0610111v2.


\bibitem{Ke85}
F. P. Kelly.: Stochastic models of computer communication systems,
Journal of the Royal Statistical Society B47, 379-395, (1985).

\bibitem{LMM07}
Y. Lu, C. M$\acute{e}$asson and A. Montanari.: TP decoding,
http://arxiv.org/PS\_cache/arxiv/pdf/0710/0710.0564v1.

\bibitem{Ly89}
R. Lyons, The Ising model and percolation on trees and treelike
graphs, Comm. Math. Phys., 125(2), (1989), 337-353.

\bibitem{MS08}
E. Mossel and A. Sly.: Rapid mixing of gibbs sampling on graphs that
are sparse on average, To Appear in SODA 2008.


\bibitem{NT07}
C. Nair and P. Tetali.: The correlation decay (CD) tree and strong
spatial mixing in multi-spin systems, Preprint on
http://front.math.ucdavis.edu/math.PR/0701494.


\bibitem{We05}
D. Weitz.: Combinatorial cirteria for uniqueness of Gibbs measures,
Random Structures and Algorithms 27, 445-475, (2005).

\bibitem{We06}
D. Weitz.: Counting indpendent sets up to the tree threshold,
Proceedings of the 38th annual ACM symposium on Theory of
computing(STOC), 140-149, (2006).

\bibitem{ZLB09}
J. Zhang, H. Liang and F. Bai.: Approximating partition functions of
two-state spin systems, submitted.
\end{thebibliography}
\end{document}